# Technical Report

25 March 2019

*Dataset of an EEG-based BCI experiment in Virtual Reality and on a Personal Computer*

~

G. Cattan, A. Andreev, P. L. C. Rodrigues, M. Congedo

GIPSA-lab, CNRS, University Grenoble-Alpes, Grenoble INP.
Address : GIPSA-lab, 11 rue des Mathématiques, Grenoble Campus BP46, F-38402, France

**Introduction**

In this work, we focus on the so-called P300 Brain-Computer Interface (BCI), a stable and accurate BCI paradigm relying on the recognition of a positive potential occurring in the EEG peaking 240 to 600 ms after stimulation onset. We implemented such a BCI on an ordinary and affordable smartphone-based head-mounted VR device and compared the user experience and the performance of the BCI on VR with a traditional BCI running on a Personal Computer (PC). An example of application of this dataset can be seen in (1).

**Participant**

A total of 21 volunteers participated in the experiment (7 females), with mean (sd) age 26.38 (5.78) and median age 26. 18 subjects were between 19 and 28 years old. Three subjects with age 33, 38 and 44 were outside this range. All participants provided written informed consent confirming the notification of the experimental process, the data management procedures and the right to withdraw from the experiment at any moment. The study was approved by the Ethical Committee of the University of Grenoble Alpes (Comité d'Ethique pour la Recherche Non-Interventionnelle).

**Material**

*VR*

We have chosen the usage of passive HMD, consisting of a mask with no electronics and a regular smartphone. Among these masks, the VRElegiant headset (Elegiant, Austin, US) is affordable, comfortable and adapts to a wide range of smartphones (**Figure 1b**). The VRElegiant was coupled with a Huawei Ascend mate 7 (Huawei, Seoul, South Korea). The mate 7 (**Figure 1a**) was a middle-range smartphone, affordable for the general public. It also has a large screen (1920 x 1080), which makes it very interesting to improve the immersion feeling in VR.

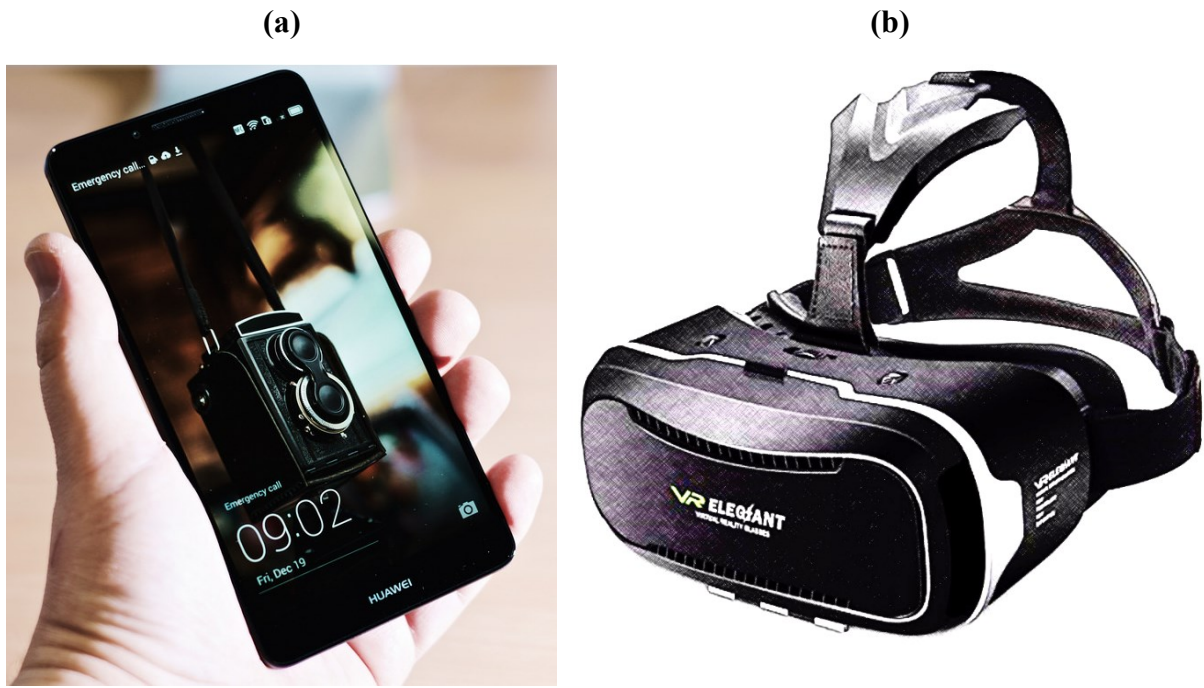

**Figure 1.** (a) Huawei Mate 7 (Huawei, Shenzhen, Chine) and (b) VRElegiant (Elegiant, Austin, US). From: www.flickr.com and https://photomania.net.

*EEG*

EEG signals were acquired by means of a standard research grade amplifier (g.USBamp, g.tec, Schiedlberg, Austria) and the EC20 cap equipped with 16 wet electrodes (EasyCap, Herrsching am Ammersee, Germany), placed according to the 10-10 international system (**Figure 2**). The locations of the electrodes were FP1, FP2, FC5, FC6, FZ, T7, CZ, T8, P7, P3, PZ, P4, P8, O1, Oz, and O2. The reference was placed on the right earlobe and the ground at the AFZ scalp location. The amplifier was linked by USB connection to the PC where the data were acquired by means of the software OpenVibe (2,3). Data were acquired with no digital filter applied and a sampling frequency of 512 samples per second. For ensuing analyses, the application tagged the EEG using USB. The tag were sent by the application to the amplifier through the USB port of the PC or smartphone. It was then recorded along with the EEG signal as a supplementary channel. The tagging process was different under PC and VR because 1) a mini-USB to USB adaptor was necessary for the smartphone (VR), 2) two different serial port communication libraries were used on VR and PC (4,5) and 3) the smartphone screen was turned 90° in VR. This results in a different mean tagging latency as explained in (6). The mean tagging latency were around 38ms in PC and 117ms in VR. These mean latencies (in PC and VR) should be used to correct ERPs by shifting them respectively in PC and VR condition (7).

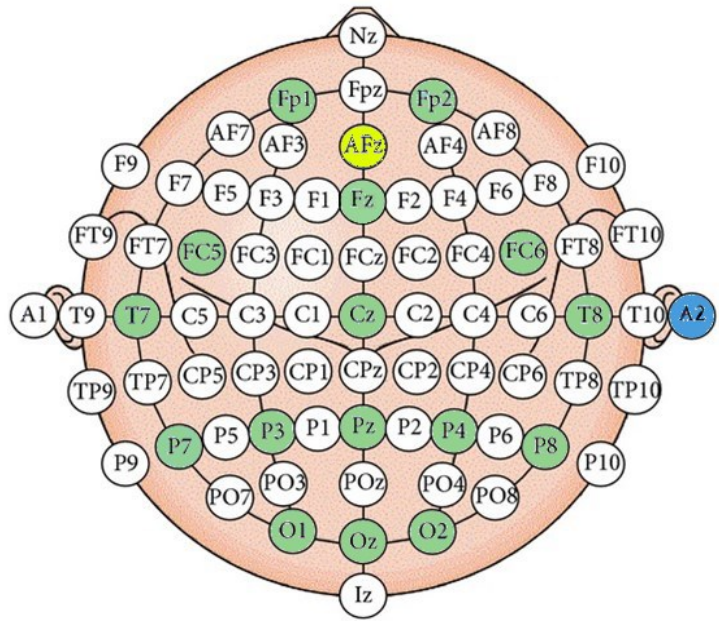

**Figure 2.** In green, the 16 electrodes placed according to the 10-10 international system. We used AFZ (in yellow) as ground and A2 (in blue) as a reference.

**Procedure**

In order to compare the use of BCI with HMD (VR) and without HMD (PC), we developed a simple P300 interface consisting of a six by six matrix of white flashing crosses. The task of the subject was to focus on a red-squared target (**Figure 3**). The user interface was identical for the PC and VR conditions. It was implemented within the Unity engine (Unity, San Francisco, US) before being exported to the PC and VR platforms thanks to the engine. In this way, we ensure that the visual stimulation is the same in the two experimental conditions.

The experiment was composed of two sessions. One session ran under the PC condition and the other under the VR condition. The order of the session was randomized for all subjects. Each session comprised 12 blocks of five repetitions. All the repetitions within a block have the same target. A repetition consisted in 12 flashes of groups of six symbols chosen in such a way that after each repetition each symbol has flashed exactly two times (8,9). Thus in each repetition the target symbol flashes two times, whereas the remaining ten flashes do not concern the target (non-target). The EEG signal was tagged corresponding to each flash.

After each repetition a random feedback was given to the subject in the form of the BCI item selection. The feedback was 'correct' if the selected symbol was the target, 'incorrect' otherwise. The probability of the feedback to be correct was drawn randomly from a uniform distribution with expectation 70%. The use of a random feedback ensures that the performance of a participant does not depend on the feedback, avoiding confounding effects due to inter-subject variability, for instance, the perceived confidence or frustration in operating the BCI, which may affect the actual performance and concentration of the participants.

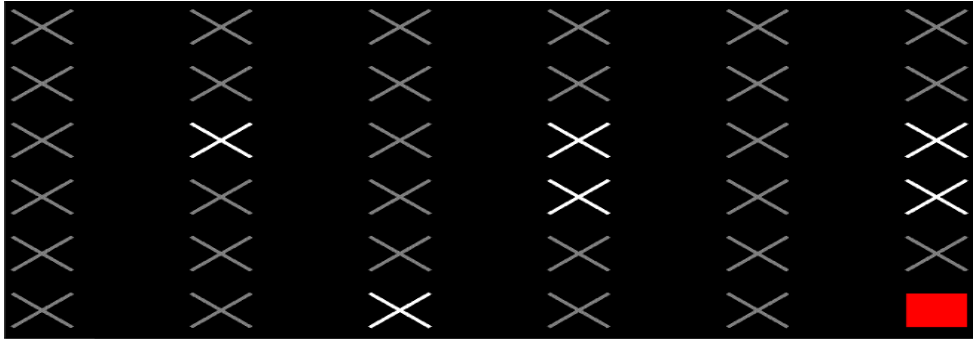

**Figure 3.** User interface at the moment when a group of six symbols is flashing. The red-square symbol is the target. The cross signs the non-target.

A pilot experiment showed that the Inertial Measurement Unit (IMU) of the smartphone sometimes accumulated an unexpected amount of drift, causing the virtual world slowly moving around the subject. Therefore in the experiment the IMU was deactivated, thus the application was always fixed in front of the subject's eyes.

**Questionnaire**

At the end of the experiment, each subjects filled in a questionnaire. This questionnaire was composed of ten questions presented in **Table 1** with their corresponding variable name. **Table 2** presents three other variables we computed on the basis of the value of these ten first variables. When the question was open such as "How many hour did you play First Player Shooter a week?", the authors associated a categorical variable to this question and created the levels.

| Number | Question | Variable name in dataset |
|---|---|---|
| 1 | Evaluate your tiredness before the experiment on a scale from 0 to 10 where 0 is 'no fatigue' | FatigueBefore |
| 2 | Evaluate your tiredness after the experiment on a scale from 0 to 10 where 0 is 'no fatigue' | FatigueAfter |
| 3 | Did you feel a sensation of discomfort? | DoesParticipantFeelDiscomfort (1 for a positive answer to question 3, 0 elsewhere) |
| 4 | Did you prefer the PC or VR session (answer : PC, VR, SAME) | DoesParticipantPreferVR |
| 5 | Evaluate your sensation of control under PC on a scale from 0 to 10 (0 = 'no control') | SensationOfControlInPC |
| 6 | Evaluate your sensation of control under VR on a scale from 0 to 10 (0 = 'no control') | SensationOfControlInVR |
| 7 | How many hours a week do you play video games? | XP_VG (1 for none; 2 for occasional; 3 for regular) |
| 8 | How many hours a week do you play First Player Shooter? | XP_FPS (1 for none; 2 for occasional; 3 for regular) |
| 9 | Have you ever experimented Virtual Reality? If yes, how many times? | XP_VR( 1 for none; 2 for occasional and 3 for repetitive experience in VR) |
| 10 | Please circle your gender: Male - Female. | IsMan (0 for female; 1 for male) |
| 11 | How old are you? | Age |

**Table 1.** Questionnaire

| Variable name in dataset | Description |
|---|---|
| FatigueDiff | FatigueAfter - FatigueBefore |
| SensationOfControlPreference | 1 if the sensation of control under PC was greater than the sensation of control under VR, 2 if vice versa. |
| IsVRSessionFirst | 1 if VR session was presented first and 0 otherwise |

**Table 2.** Description of factors and their levels

**Organization of the dataset**

For each subject we provide two *mat* (and *csv*) files (Mathworks, Natick, USA) containing the complete recording of the sessions in the two experimental conditions (VR and PC). Each file is a 2D-matrix where the rows contain the observations at each time sample. Columns 2 to 17 contain the recordings on each of the 16 EEG electrodes. The first column of the matrix represents the timestamp of each observation and column 18 contains the experimental events. The rows in column 18 (Event) are filled with zeros, except at the timestamp corresponding to the beginning of an event, when the row gets one of the following values:

- 102 for the end of a repetition.
- 100 for the onset of a new block.
- 20-25 and 40-45 when a group which does not contain the target flashes. The twelve groups are separated in six rows and six columns, in such way that a symbol is included in exactly one row and one column (9). Note that the naming of rows and columns do not refers to the physical rows and columns in the matrix, although it was the case in the first implementation of the protocol (10). The first digit of the values indicates whether the group is a "row" (digit *2*) or a "column" (digit *4*). The second digit indicates the number of the flashed row or column in the range [0, 5]. Note that the groups are randomized between the repetitions thus a physical symbol in the matrix does not corresponds to the same row or column.
- 60-65 and 80-85 when a group containing the target flashes. The first digit of the values indicates whether the group is a row (digit *6*) or a column (digit *8*). The second digit indicates the number of the flashed row or column in the range [0, 5].

For ease of use, we provide columns 19 to 22, which are filled with zeros, except at the timestamp corresponding to the following events:

- Column 19 (IsTarget) contains one when a group containing the target flashes.
- Column 20 (IsNonTarget) contains one when a group which do not contains a target flashes.
- Column 21 (IsStartOfNewBlock) contains a positive integer in the range [1, 12]. Values correspond to the number of the started block.
- Column 22 (EndOfRepetitionNumber) contains a positive integer in the range [1, 5]. Values corresponds to the number of achieved repetition for a same target.

The *Header.mat* (or *Header.csv*) file contains the column names, sorted by ascending column number, including the name of the EEG channels we used. We also provide a *Questionnaire.mat* (and *Questionnaire.csv*) file which contains, for each subject, the value of the 14 variables presented in **Table 1** and **Table 2**. Note that the questionnaire also includes the demographic variables, that is, the genre and age of the subjects. The names of the variable within the *Questionnaire.mat* (and *Questionnaire.csv*) file are reported in the *Questionnaire_header.mat* (and *Questionnaire_header.csv*) file.

We supply also an open-source Python code example (1) using the analysis framework MNE (11,12) and MOABB (13,14), a comprehensive benchmark framework for testing popular BCI classification algorithms. This python code is a classification example of the P300 using epochs of signal with duration one second. The performance of the classifier is evaluated in PC and VR. A complete analysis of this dataset is been submitted for publication at the time of writing.

**References**

1. Coelho Rodrigues PL. EEG-Virtual-Reality [Internet]. Grenoble: GIPSA-lab; 2018. Available from: https://github.com/plcrodrigues/py.VR.EEG.2018-GIPSA

2. Renard Y, Lotte F, Gibert G, Congedo M, Maby E, Delannoy V, et al. OpenViBE: An Open-Source Software Platform to Design, Test, and Use Brain–Computer Interfaces in Real and Virtual Environments. Presence Teleoperators Virtual Environ. 2010 Feb 1;19(1):35–53.

3. Arrouët C, Congedo M, Marvie J-E, Lamarche F, Lécuyer A, Arnaldi B. Open-ViBE: A Three Dimensional Platform for Real-Time Neuroscience. J Neurother. 2005 Jul 8;9(1):3–25.

4. Mandal MK. C++ Library for Serial Communication with Arduino [Internet]. 2016. Available from: https://github.com/manashmndl/SerialPort

5. Wakerly M. Usb-Serial-For-Android [Internet]. 2012. Available from: https://github.com/mik3y/usb-serial-for-android

6. Cattan G, Andreev A, Maureille B, Congedo M. Analysis of tagging latency when comparing event-related potentials [Internet]. Gipsa-Lab ; IHMTEK; 2018 Dec. Available from: https://hal.archives-ouvertes.fr/hal-01947551

7. Cattan G, Andreev A, Cesar M, Marco C. Toward Deploying Brain Computer Interfaces to Virtual Reality Devices Running on Ordinary Smartphones. submitted.

8. Andreev A, Barachant A, Lotte F, Congedo M. Recreational Applications of OpenViBE: Brain Invaders and Use-the-Force [Internet]. Vol. chap. 14. John Wiley ; Sons; 2016 [cited 2017 Aug 3]. Available from: https://hal.archives-ouvertes.fr/hal-01366873/document

9. Congedo M, Goyat M, Tarrin N, Ionescu G, Varnet L, Rivet B, et al. "Brain Invaders": a prototype of an open-source P300- based video game working with the OpenViBE platform. In: 5th International Brain-Computer Interface Conference 2011 (BCI 2011) [Internet]. 2011 [cited 2017 Aug 1]. p. 280–3. Available from: https://hal.archives-ouvertes.fr/hal-00641412/document

10. Guger C, Daban S, Sellers E, Holzner C, Krausz G, Carabalona R, et al. How many people are able to control a P300-based brain-computer interface (BCI)? Neurosci Lett. 2009 Oct 2;462(1):94–8.

11. Gramfort A, Luessi M, Larson E, Engemann DA, Strohmeier D, Brodbeck C, et al. MNE software for processing MEG and EEG data. NeuroImage. 2014 Feb 1;86:446–60.

12. Gramfort A, Luessi M, Larson E, Engemann DA, Strohmeier D, Brodbeck C, et al. MEG and EEG data analysis with MNE-Python. Front Neurosci [Internet]. 2013;7. Available from: https://www.frontiersin.org/articles/10.3389/fnins.2013.00267/full

13. Barachant A. Mother of All BCI Benchmarks. [Internet]. NeuroTechX; 2017. Available from: https://github.com/NeuroTechX/moabb

14. Jayaram V, Barachant A. MOABB: trustworthy algorithm benchmarking for BCIs. J Neural Eng. 2018 Sep 25;15(6):066011.